\documentclass[12pt]{article}

\begin{document}

\title{On the logic of quantum physics\\
and the concept of the time}

\author{Luigi Foschini\\
{\small Institute TeSRE -- CNR, Via Gobetti 101, I-40129, Bologna (Italy)}\\
{\small E-mail: \texttt{foschini@tesre.bo.cnr.it}}}

\maketitle

\begin{abstract}
The logic--linguistic structure of quantum physics is analysed. 
The role of formal systems and interpretations in the representation of nature is
investigated. The problems of decidability, completeness, and consistency can affect
quantum physics in different ways. Bohr's complementarity is of great interest,
because it is a contradictory proposition. We shall see that the 
flowing of time prevents the birth of contradictions in nature, 
because it makes a cut between two different, but complementary 
aspects of the reality.
\vskip 12pt
\noindent PACS: 03.65.-w Quantum mechanics; 03.65.Bz Foundations, 
theory of measurement, miscellaneous theories; 02.10.By Logic and 
foundations.\\
\end{abstract}

\section{Introduction}
The foundations of physics, the quantum logic, the interpretation of 
quantum mechanics: since the beginning, quantum theories gave rise to 
a broad variety of studies not properly defined studies devoted to 
settle the properties of this new branch of physics. Studies about 
quantum logic started in 1936, with the work of Birkhoff and von 
Neumann \cite{BIRKHOFF} and continued with many solutions, such as 
three-valued logics or unsharp approaches \cite{BELTRAMETTI}, 
\cite{JAMMER}.  However, several authors, even though with some 
differences, point out that quantum mechanics does not require a new 
non-classical or unconventional logic \cite{GRIFFITHS}, \cite{OMNES}, 
\cite{ORLOV}.  Studies about the interpretation of quantum mechanics 
are quite plethoric \cite{JAMMER}, \cite{WHEELER83}.  A widely 
accepted interpretation is the so called ``Copenhagen 
interpretation'', initially developed by Niels Bohr and Werner 
Heisenberg, among others.  However, the debate is still alive.  There 
is no fundamental disagreement among physicists on how to use the 
theory for practical purposes, such as to compute energy levels, 
transition rates, etc.  \cite{PERES95}.  Problems arise on the 
concepts of quantum theory.

Many authors asked for a new language, different from the everyday 
one.  Bohr stressed the difficulties of all means of expressing 
ourselves: human beings are ``suspended in language'', as he used to 
say (see \cite{PETERSEN}).  People strive to communicate experiences 
and ideas to others and to extend their field of descriptions, but 
this should be done in a way so that messages do not become too 
ambiguous.  As far as the definition of experiment is concerned, Bohr 
said that it is a situation where it is possible to tell others what 
has been done and what has been learned and thus, observations results 
should be expressed in an unambiguous language \cite{BOHR49}.

The problem of the language of physics is not trivial.  It could seem 
a negligible detail, because traditional philosophy regarded language 
as something secondary with respect to empirical reality.  Still today 
there are many physicists believing that elementary particles are as 
real as an apple or a grapefruit can be (see \cite{HORGAN}); Asher 
Peres emphasizes that experiments do not occur in a Hilbert space, but 
in a laboratory \cite{PERES95}.  Many times one makes the mistake to 
search the meaning of a word or a symbol as something coexisting with 
the sign: there is a (con)fusion between \emph{substance} and 
\emph{substantive} \cite{WITT}.

The distinction between the mathematical symbol and the physical 
object is of paramount importance, because it allows to define what 
physics is.  Mathematics is a language, invented by mankind, that we 
use to speak about nature and not \emph{the} nature itself.  Niels 
Bohr said that physics is simply what we can \emph{say} about nature, 
not a way to find out how nature \emph{is} (see \cite{PETERSEN}). 
There is no doubt about the existence of a world, but we must consider 
that no experience can be understood or communicated without a 
logic--linguistic frame. Both scholars and scientists elaborate ideas 
and report on the results. To make this they need of, at least, two 
languages: mothertongue and mathematics. We must learn to use them as 
any other laboratory instrument.

In this paper the reader does not find something of practical or 
predictive.  We would like, in agreement with Bohr, to say something 
about nature by using the language of mathematics.  A grain of 
knowledge, hoping that it will be a \emph{granum salis}. We have never seen something
like the elaboration exposed in this paper and even though 
some things in the following sections may appear trivial, we think it 
is necessary to review them, in order to fix some definitions and 
notations at least. We hope it will be useful to see some old concepts in a
different way. Moreover, we would like to point out a new possible definition of the
concept of the time.

\section{Formal systems}
Before going on, let us recall some notions of formal systems.  It is 
possible to distinguish between \emph{formal logistic systems} and 
\emph{interpreted languages}.  In the first case, syntactical rules 
are sufficient to determine the system, in the second one semantic 
rules are also necessary.  Generally speaking, a formal system is 
constituted by \cite{KLEENE52}, \cite{KLEENE67}, \cite{LOLLI}:

\begin{enumerate}
	\item a complete list of formal symbols or signs with which the 
	expressions are formed;

	\item an explicit, or also recursive, definition of what a 
	well-defined formula is and what a term is;

	\item a numerable set of well-defined formulas taken as axioms;

	\item some metamathematical definitions, called rules of 
	inference, which allow the transformation of formulas in other 
	formulas or show what is an immediate consequence of a formula;

	\item a list of formulas which can be derived with a finite number 
	of applications of the rules of inference;

	\item a list of sentences that allow us to shorten long 
	expressions;

	\item a list of sentences about the language that explicitly shows 
	the denotation properties and, particularly, an indication of what 
	expressions they denote and the circumstances in which an 
	expression denotes a particular object.
\end{enumerate}

Let us indicate the first type of formal system (formal logistic 
system or calculus) with the symbol $FS(\vdash)$, where $\vdash$ is 
the well known symbol for formal deduction.  $FS(\vdash)$ has the 
properties from 1 to 6 indicated above.  We then indicate a formal 
system with semantics, or an interpreted language, with the symbol 
$FS(\vdash, \models)$, where $\models$ is the symbol for the semantic 
deduction.  $FS(\vdash, \models)$ has the proprieties from 1 to 7 
indicated above.  The statement n.~7 is present for semantic systems 
only.

Now it is possible to construct several languages for various 
purposes, from everyday talking to science languages \cite{CARNAP58}, 
\cite{CARNAP63}.  It should be pointed out that there is a small 
difference between the list of mathematical logic and the alphabet of 
a language: in the second case, with a given interpretation, many of 
the formal symbols correspond to entire words and phrases rather than 
to a single letter \cite{KLEENE52}.  The postulates and axioms for 
mathematics can be found in \cite{KLEENE52}.

Statement n.~7, which makes the distinction between calculus and its 
interpretation, leads us to make a distinction between two types of 
truth, logic truth and factual one \cite{CARNAP63}.  Truth in 
mathematics (a formal logistic system) is indipendent from the objects 
of the world and must be valid for every possible interpretation.  
Truth in physics (an interpreted language) is considered factual: a 
sentence is true when it corresponds to reality.  This distinction is 
extremely important: in 1931, Tarski showed that it is possible to 
define truth only in purely formal languages \cite{TARSKI1}.  Truth in 
an interpreted language cannot be defined inside the language itself.  
Tarski reported an useful example: the sentence ``it is snowing'' is 
true if and only if it is snowing.  One must look out of the window 
whether it is snowing.  This definition of truth must be extended 
according to Peirce, so that both for logic and factual truths it is 
possible to say that a proposition is true if, and only if, it is true 
what it \emph{explicitly and implicitly} declares \cite{PEIRCE2}.  
This is of paramount importance in science: it is thanks to this 
definitions that we can explore things outside our direct experience.  
For example, we can say that a black hole exists, even if we cannot 
directly observe it, because its existence follows from true 
propositions.

It is necessary to point out that mathematics is an interpreted 
language, even though the truth in this case is a logical one, 
because it is not possible to verify it with any external reality. A 
triangle is a geometric interpretation of a formal object in a 
certain formal logistic system, but it does not exist in reality. We 
can find objects with triangular shape, but not triangles. 

Therefore, being physics an interpretation of the mathematics, we can 
conclude that it is an interpretation of an interpreted language. A 
semantics of a semantics. Objects in physics does not exists in 
reality, only as approximations. In physics there is difference 
between a fluid and the water, in the same way as in mathematics there 
is difference between triangles and triangular--shaped objects. 

In conclusion, we can always speak about truth in the language (interpreted 
or not), rather than truth with direct reference to the world.

\section{Some features of formal systems}
Main features of a formal system are decidability, completeness and 
consistency. As known, at the beginning of the XX century, the German 
mathematician David Hilbert presented a list of problems, among them 
there were \cite{HILBERT}:

\begin{enumerate}
\item Decidability (\emph{Entscheidungsproblem}), i.e. a formula $A$ 
is decidable if $\vdash A \vee \vdash \neg A$.

\item Completeness (\emph{Entscheidungsdefinitheit}), i.e.  
every closed formula of the formal system is decidable.

\item Consistency (\emph{Widerspruchsfreiheit}), i.e. there are no 
contradictory formulas ($\vdash 
A \wedge \vdash \neg A$).
\end{enumerate}

Even though decidability is introductory to completeness, this problem 
was solved later, after the problems of completeness and consistency.  
Indeed, in 1931 Kurt G\"{o}del solved problems 2 and 3 \cite{GODEL}, 
while in 1936, Alan Turing \cite{TURING} and Alonzo Church 
\cite{CHURCH}, indipendently solved question 1. Probably this gives 
the misleading impression that the \emph{Entscheidungsproblem} is 
implied in G\"{o}del's work, thus without the Church--Turing thesis.

G\"{o}del demonstrated that in a formal system, like that set up by 
Whithead and Russell in their book \emph{Principia Mathematica} or 
like the Zermelo--Fraenkel--von Neumann axiom system of set theory, 
there are propositions undecidable on the basis of system's axioms, so that 
these formal systems cannot be complete (in the G\"{o}del's paper this 
is stated in Theorem VI).  This is valid for a 
wide class of formal systems, in particular for all systems that 
result from the two just mentioned with the addition of a finite 
number of axioms, \emph{i.e.} those systems that fulfil the previously 
listed statements, from 1 to 6. As G\"{o}del wrote, in a note added 28 
August 1963, the characteristic property of a formal system is that 
reasoning in them, in principle, can be completely replaced by 
mechanical devices \cite{GODEL}. Later on Tarski extended these 
results also to interpreted languages \cite{TARSKI3}.  It is worth 
noting that the undecidability of a proposition can be skipped by 
stating an axiom that defines whether that proposition is true or 
false.  However, even in such a new formal system there will be a new 
undecidable proposition.  In order to have a complete formal system, 
we should have an infinite number of axioms.

Moreover, it is necessary to note that in this case we have implicitly 
assumed that the formal system is consistent. But, as G\"{o}del himself 
showed in his famous paper, the consistency of the formal system is 
not provable in the formal system itself (this is stated in the 
Theorem XI) \cite{GODEL}.

G\"{o}del's results were later generalized by Church and Turing and 
used to solve the decision problem, which is closely connected to the 
problem of completeness, as we have already seen.  The 
\emph{Entscheidungsproblem} consists in finding a general algorithm, 
which would enable us to establish whether or not any particular 
sentence can be proved within that formal system.  Church 
\cite{CHURCH} and Turing \cite{TURING} showed that this is not 
possible. 

\section{The logical structure of physics}
Let us consider the logical structure of physics and the role of 
language.  We recognize the existence of a reality, but to study it, 
to investigate it, to communicate our results, we need a language.  
The word does not change reality, but influences our way of thinking.  
It is necessary to analyze the relationship among words and things, 
physics and reality.

When children learn to talk, they correlate an object to a word: they 
give a meaning to a word and begin to believe that language is 
meaning-based.  Only later they will know that language is 
syntax-based.  For example, inserting the word `only' into all 
possible positions in the sentence:

\begin{quotation}
	\emph{I helped Mickey Mouse eat his cheese last week.}
\end{quotation}

\noindent it is possible to see that the meaning is a function of the 
syntax. During its ``childhood'', physics was a correlation between an object 
and a mathematical symbol.  It is then possible to speak about physics 
as interpretation of mathematical symbols, just like when we speak 
about the meaning of a word.  Galileo Galilei wrote that physics is 
like a great book written in mathematical language, whose characters 
are triangles, circles, and other geometrical figures \cite{GALILEI}.  
Later on, Charles Sanders Peirce founded the logic of relations, 
noting that any deductive science establishes an isomosphism between 
the logic of nature and the logic of the language we employ to explore 
it \cite{PEIRCE1}.
 
A correlation is possible when objects can be found in the everyday 
life and everybody can see them.  It is possible to make a correlation 
between an apple and the word `apple'.  Anyone can see, touch, smell, 
eat or hear (when it falls down) an apple: we can use our five senses 
to know the apple.  Therefore, there is no doubt about the meaning of 
the word `apple'; moreover it must be stressed that there is a 
correlation only and not a (con)fusion between an apple and the word 
`apple'.

The correlation is not so clear when we speak, for example, about 
electromagnetism.  Nobody can see, touch, smell, eat or hear a radio 
wave: it is necessary to use an instrument, the radio receiver, in 
order to hear radio waves.  The use of analogies based on everyday 
life can be misleading, as known from the hydrodynamic model of 
electric current.

Things become more complicated when we speak about quantum mechanics.  
The esigence of observing a world outside our direct experience and 
with instruments that must be within human beings' reach, forces us to 
emphasize the experiment procedure \cite{PERES95}.  Only certain 
things are observable and this happens only after an irreversible act 
of amplification, such as the blackening of a grain of silver bromide 
emulsion or the triggering of a photodetector: what we choose to 
measure has an inavoidable consequence on what we shall find 
\cite{WHEELER81}.  It is possible to correlate a mathematical symbol 
only to those ``observables'', \emph{i.e.} only observables can have a 
meaning.  All other mathematical symbols or words used in quantum 
mechanics do not have any meaning.  The esigence of abstraction must 
be emphasized.

Let us consider an example: the d'Alambert wave equation, which can be 
written as following:

\begin{equation}
	\frac{\partial^{2}\psi (\mathbf{r}, t)}{\partial 
	t^{2}}-c^{2}\nabla ^{2}\psi (\mathbf{r}, t)=0
	\label{e:wave}
\end{equation}

\noindent This equation shows that there is a logical link among the symbols 
indicated.  This link is formal and then indipendent from meaning.  
Then, Eq.~(\ref{e:wave}) can be used in several branches in physics, 
according to the meaning that one gives to used symbols.  The function 
$\psi$ can be correlated, even though without a one-to-one 
correspondence, with pressure, velocity, velocity potential, 
displacement of matter points from equilibrium in a solid, electric 
field, magnetic induction or the corresponding potentials.  Moreover 
there are many other interpretations of Eq.~(\ref{e:wave}) (for 
examples, see \cite{JONSSON}).  The interpretation of other symbols 
follows as a consequence of the choice of interpretation of $\psi$.  
It should be noted that the interpretation does not directly concern 
the Eq.~(\ref{e:wave}), but through the boundary conditions or through 
the initial values.  However, it should be stressed that the wide 
application field of the wave equation does not derive from its 
physical interpretation. Starting with an interpreted wave 
equation one can obtain misleading results.  For example, one cannot 
consider pressure waves analogous to electromagnetic waves: in the 
first case there are some consequences absent in the second one 
(\emph{e.g.} the supersonic ``bang'').  The formal character of 
physics, and of any deductive science, is due to the fact that when 
deducing a theorem from the postulates, it is necessary to avoid any 
specific property of the interpretation and to use only those formal 
properties which are explicitly stated in postulates (and therefore, 
belong to every interpretation of the formal system) \cite{TARSKI2}.  
The final conclusion is that every theorem of a given deductive theory 
is satisfied by any interpretation of the formal system of this theory 
\cite{TARSKI2}.

Let us now consider the Schr\"{o}dinger wave equation:

\begin{equation}
	i\hbar\frac{\partial\psi (\mathbf{r}, t)}{\partial 
	t}+\frac{\hbar^{2}}{2m}\nabla ^{2}\psi (\mathbf{r}, t)-U\psi 
	(\mathbf{r}, t)=0
	\label{e:qwave}
\end{equation}

\noindent It must be noted that $\psi$, at this stage, has no meaning and it is 
not possible to find anyone.  As known, it is necessary to make 
further operations on $\psi$.  For example, the probability $dP$ to 
find an electron in a volume $dV$ centered in the point $\mathbf{r}$ 
is:

\begin{equation}
	dP=|\psi(\mathbf{r})|^{2}dV
	\label{e:prob}
\end{equation}

There is no strict correlation between the symbol and the object.  It 
should be underlined that experiments are always macroscopic, because 
they must be readable by a human being.  We cannot see an electron 
directly, but we can observe some effect on some instrument.  This 
does not involve any particular logic or physical law, but a strong 
attention to the interpretation.

The question of the correlation between a physical object and a 
mathematical symbol in quantum theory gave rise to a wide debate.  The 
most famous one occurred between Albert Einstein and Niels Bohr and is 
synthetized in the EPR paradox \cite{BOHR35}, \cite{EPR}.  Already in 
the initial words of the Einstein, Podolski and Rosen's paper 
\cite{EPR} it is possible to see the authors' conception the physics: 
each element of reality corresponds to an element in the theory.  This 
type of physics is strongly based on the meaning of a word or a 
symbol.  Bohr's reply shows that this conception is ambiguous and 
then, it can generate paradoxes \cite{BOHR35}.  Bohr underlined that 
quantum mechanics forces us to a radical revision of the traditional 
idea of an absolute physical reality, as Einstein's relativity led us 
to abandone the idea of an absolute space.  In each experiment, owing 
to the impossibility of discrimination, between who is the observer 
and what is observed, the experiment became simply a situation where 
we can tell others what we have done (see also \cite{BOHR28}, 
\cite{BOHR49}).  Experimental procedures must be strictly specified in 
order to minimize ambiguities.

Physics is then an interpreted language that we use in order to say 
something about nature (see \cite{PETERSEN}) and the syntax of this 
language is isomorphic to the logic of nature.  Physics is 
constructed, as every language, following the rules listed in 
Section~2.  We can then write that classical physics is a semantic 
system, $FS(\vdash, \models)$, where only the experiment can tell if a 
sentence is true.  The system must be consistent, that is each 
provable formula must be valid.  It is possible to write that a 
formula $A$ is said to be true when:

\begin{equation}
	if\ \vdash A\ then\ \models A
	\label{e:cons}
\end{equation}

\noindent that is the \emph{consistency condition}.  
Eq.~(\ref{e:cons}) says that this conditional statement is an 
implication, that is $\models A$ is true under all conditions for 
which $\vdash A$ is provable.  It is also possible to say that the 
provability of $\vdash A$ ``forces'' $\models A$ to be true.

On the other hand, the \emph{completeness condition} says that each 
valid formula is provable.  By using symbols:

\begin{equation}
	if\ \models A\ then\ \vdash A
	\label{e:comp}
\end{equation}

Realists think that only Eq.~(\ref{e:comp}) is valid in physics.  When 
the semantic deduction is absent, as in quantum mechanics, realists 
force the existence of $\models A$, according to their own logic.  In 
doing this, realists find paradoxes, that, however, are in assuming 
their own logic as the semantic deduction isomorphic to the one 
of the nature.

Indeed, in quantum mechanics, the main problem is the measurement, 
that is to find a meaning for mathematical symbols.  First of all, it 
is worth noting that the existence of the Heisenberg's indeterminacy 
principle restricts the variables that can be measured simultaneously 
\cite{HEISENBERG}.  It should be noted that even in classical physics 
it is not possible to measure two variables, or more, 
\emph{simultaneously}, owing to experimental uncertainties; 
nevertheless, taking into account the smallness of these uncertainties 
with respect to macroscopic variables, it is possible to assume that 
the measurement are \emph{simultaneous}.  In quantum mechanics, in 
addition to experimental uncertainties, there is also Heisenberg's 
principle, that allows to measure canonically conjugate quantities 
simultaneously only with a characteristic indeterminacy.

This feature called for a new concept of measurement, developed and 
analyzed by several authors during this century \cite{BOHR49}, 
\cite{BUSCH}, \cite{PERES95}, \cite{NEUMANN}; see also in 
\cite{WHEELER83}.  However, Bohr \cite{BOHR49} (see also 
\cite{PETERSEN}) and, later, Peres \cite{PERES95}, wrote against the 
existence of a ``quantum measurement''; Bohr designed, in a 
semi-serious style, a series of devices that should serve to make 
these measurements \cite{BOHR49}.

It is necessary to stress that an experiment is always macroscopic, in 
order to be accessible to a human being.  We never directly observe a 
microscopic variable: nobody has very \emph{seen} a single photon, but 
only some \emph{macroscopic irreversible effects}.  In order to 
explain these effects, we use quantum mechanics, from which we can 
reconstruct a visualization of a process outside our direct 
experience.  

The postulates of quantum mechanics stated by von Neumann 
\cite{NEUMANN} are often considered as providing an interpretation in 
themselves, but it is necessary to restrict the notion of 
interpretation and to give a meaning only to those words or symbols 
which have referents, that are always macroscopic in order to be 
reachable by human beings.  These axioms and postulates are widely 
dealt in those books and we refer to them for readers interested in 
further investigations.

Having a formal system, we can apply G\"{o}del's procedure and make 
the arithmetization of the quantum theory.  To do this, we must assign 
an integer to a sequence of signs or to a sequence of sequences of 
signs.  For example, it is possible to use the number 5 instead of the 
word `not' or 9 instead of `for all', and so on.  Notions such as 
`formula' and `provable formula' can now be represented by integers or 
sequences of integers.  This method is today commonly used: in a 
computer we represent all concepts by using sequences of 0 and 1 only.  
Indeed, Turing used G\"{o}del's numbers in order to construct his 
famous machine, that is the logical basis of modern computers 
\cite{TURING}.

Now it is not necessary to repeat the proof of incompleteness: it is 
sufficient to have shown that quantum theory is a formal system and 
then incompleteness logically follows, as shown by G\"{o}del. The extension
of incompleteness to physics is quite straightforward, as underlined by
Svozil by means of other techniques \cite{SVOZIL}.

\section{Complementarity, Indeterminacy, Contradictions}
Many scientists tried to see in physics reminiscences or analogies 
with G\"{o}del's theorems: Peres and Zurek looked for analogies with 
quantum theory \cite{PERES85}, \cite{PERZUR}; Svozil extended 
incompleteness theorems to physics by using Turing's proof 
\cite{SVOZIL}; Komar investigated the quantum field theory 
\cite{KOMAR}; Casti examined the interconnections between chaos and 
G\"{o}del's theorem \cite{CASTI} and last, but not least, Chaitin 
analysed incompleteness in information theory \cite{CHAITIN}.  It is 
worth to note the works by Chaitin \cite{CHAITIN} and Svozil's 
\cite{SVOZIL}: they suggested that incompleteness was natural and 
widespread rather than pathological and unusual. Particularly, Svozil
extended undecidability to physics, by means of the Turing's proof
\cite{SVOZIL}. 

Scientists often connected undecidability with Heisenberg's indeterminacy 
(uncertainty) in physics.  However, undecidability does not mean 
uncertainty, even though they can be linked somehow.  It is possible 
to demonstrate that uncertainty cannot be undecidable: for example, 
let us consider the uncertainty relation for position and momentum:

\begin{equation}
	\Delta x \Delta p \geq \frac{\hbar}{2}
	\label{e:uncert}
\end{equation}

\noindent This proposition holds, while the negation of the 
uncertainty principle, \emph{i.e.}:

\begin{equation}
	\Delta x \Delta p = 0
	\label{e:uncert2}
\end{equation}

\noindent does not hold.  We recall that, a formula $A$ is said to be 
undecidable when it is not possible to prove nor $A$ neither $\neg A$.  
For uncertainty we can prove Eq.~(\ref{e:uncert}), but we 
cannot prove Eq.~(\ref{e:uncert2}), where 
Eq.~(\ref{e:uncert2})$= \neg$~Eq.~(\ref{e:uncert}).  \emph{Then, 
uncertainty is not an undecidable proposition}.

Other scientists believe that physics cannot be subjected to 
G\"{o}del's theorems because it has variables of a higher order, such 
as experimental data, but Tarski has already shown that this is not 
sufficient \cite{TARSKI3}. Morevoer, Svozil recently showed that undecidability is
present in physics, so that it cannot be complete \cite{SVOZIL}.

We want now to explore the consistency of quantum physics and a particular feature,
Bohr's complementarity, taking into account the logic--linguistic analysis 
carried out untill now. 
It is necessary to analyze also some implicit postulates, holding in 
physics: for example, all propositions must be consistent. As a 
consequence of G\"{o}del's Theorem XI (it is not possible to prove the 
consistency of a formal system within the system), 
condition~(\ref{e:cons}) must be assumed as an axiom:

\begin{quotation}
	Axiom 1: $if\ \vdash A\ then\ \models A$
\end{quotation}

Another implicit axiom is that each event is unique (\emph{i.e.}, 
there is only a single world).  As we know from Heisenberg's 
indeterminacy, it is not possible to measure simultaneously position 
and momentum, while Bohr's complementarity tells us that, even though 
particle and wave behaviour are mutually exclusive, they are also 
complementary \cite{BOHR28}.  There is no well-defined proposition 
that states what complementarity is.  Bohr wrote:

\begin{quotation}
	The very nature of the quantum theory thus forces us to regard the 
	space-time co-ordination and the claim of causality, the union of 
	which characterizes the classical theories, as complementary but 
	exclusive features of the description, symbolizing the 
	idealization of observation and definition respectively.  [\ldots] 
	we learn from quantum theory that the appropriateness of our usual 
	causal space-time description depends entirely upon the small 
	value of the quantum of action as compared to the actions involved 
	in ordinary sense perceptions \cite{BOHR28}.
\end{quotation}

We must stress the importance of time in this quotation, that Bohr 
recalled also in his reply to Einstein, Podolsky and Rosen 
\cite{BOHR35}.  The implicit postulate of uniqueness of each event 
must be reformulated in:

\begin{quotation}
	Axiom 2: \emph{There is a single world at a time t}.
\end{quotation}

Already at first sight, Bohr's complementarity can be seen as a 
contradictory proposition because, for example, an electron cannot 
behave simultaneously as a particle or as a wave.  In the famous 
experiment of the double slit, we can observe interference (wave) if 
there is no specification of which path (particle).  But if the path 
is specified, then the interference will disappears.  During these 
last years, there was a vivid debate around complementarity: it 
started with some experiments carried out by Scully \emph{et al.} 
\cite{SCULLY}, where they claimed to have observed the simultaneous 
behaviour of waves and particles, avoiding the indeterminacy 
principle.  This episode could be dramatic for physics, because a 
simultaneous observation of wave and particle behaviour would 
contradict the implicit Axiom 2.  Some years later, Storey \emph{et 
al.} replied that the Heisenberg's principle is valid and then Scully 
\emph{et al.} were wrong \cite{STOREY94}.  This was followed by a 
letter of Englert \emph{et al.} \cite{ENGLERT}, to which Storey 
\emph{et al.} soon reply \cite{STOREY95}.  Moreover, Wiseman and 
Harrison wrote a letter in which they claimed that both groups were 
right \cite{WISEMAN}.  

It is worth to note that in all these 
discussions no adequate attention was given to time.  The 
indeterminacy principle says that, at a given time $t$, we can 
observe, with sufficient precision, one canonically conjugate variable 
only: position or momentum, particle or wave.  When we prepare an 
experiment, we choose what we want to observe, particle or wave. This is
equal to select a particular interpretation (semantic deduction) for the
formal system. Let us consider these two propositions:

\begin{itemize}
	\item $A$ = ``an electron behave like a wave''

	\item $B$ = ``an electron behave like a particle''
\end{itemize}

\noindent which are complementary, as known.  Owing to the fact that 
at a time $t$ we must have a single world, it is possible to write 
that:

\begin{equation}
	B = \neg A \ (or \ A = \neg B)
	\label{e:comple}
\end{equation}

Let us suppose that we set up an experiment to observe, say, the 
particle behaviour of an electron.  So $B$ is true, while $A = \neg B$ 
is false.  \emph{In a second time}, we can set up a new experiment to 
detect the wave behaviour of the electron.  In this case, $A$ is true, 
while $B = \neg A$ is false.  If we want to construct a formal system 
for quantum mechanics we should select one of these behaviours, but it 
is not possible to set up an experiment that allows to decide whether 
$A$ or $B$ is always true.  We have that both can be true and false: 
this is a contradiction.

Time is the only thing that prevents the birth of a contradiction in 
nature.  The flowing of time makes the existence of two different 
aspects possible, avoiding the contradiction of a simultaneous 
existence (intuitively we could see time as a ``cut'' in space).  
Neglecting time in quantum mechanics could then lead to contradictory 
propositions, because both $A$ and $\neg A$ (or $B$ and $\neg B$) 
simultaneously hold and this contradicts Axiom 2.  Time creates a cut 
between two different, but complementary aspects of reality.  It is 
also possible to deduce that we cannot unificate physics under a 
single description: we need both waves and particles.

\section{Time in quantum mechanics}
``How much time'', ``It took quite some time'', ``There is plenty of 
time''.  These are some examples only of our common way to think about 
time: an interval between two instants.  In physics as well time is 
seen as an interval.  Asher Peres stressed that the measurement of 
time is the observation of a dynamical variable, which law of motion 
is known and it is uniform and constant in time \cite{PERES95}.  There is a 
kind of self--reference in this definition and nothing is said about 
time.  On the other hand, time is considered simply as a parameter 
and, according to this, the above definition is completely 
satisfactory.  Moreover, time is sometimes neglected (e.g.  steady 
state phenomena), which is useful to understand some physical 
concepts.

However, when we deal with quantum mechanics the problem of the time 
explodes in all its complexity.  In classical mechanics (hamiltonian 
formulation) the dynamical state of a physical system is described by 
a point in a phase space, that is we have to know position $q$ and 
momentum $p$ at a given time $t$.  Even though it can appear a 
sophism, it is not possible, strictly speaking, to know simultaneously 
$q$ and $p$ of any object.  However, in classical physics, we may 
neglect variations during the lapse of time between the measurement of 
$q$ and $p$, because the quantum of action is so small when compared 
to macroscopic actions.  It is very interesting to note how Sommerfeld 
stressed this question when he wrote about the Hamilton's principle of 
least action: the trajectory points $q$ and $q+\delta q$ are 
considered \emph{at the same time instant} \cite{SOMMERFELD}.

In quantum mechanics this approximation is not valid, because actions 
are comparable with the quantum of action.  The hamiltonian formalism 
is not anymore a useful language to investigate nature and, as known, 
it was necessary to settle quantum mechanics.  

The impossibility to neglect time in quantum mechanics is well described 
by the Heisenberg's principle of indeterminacy. Nevertheless, the role of 
the time in quantum indeterminacy is often neglected. In the history 
of physics, we can often find authors which claimed to 
have found a way to avoid the obstacle of indeterminacy. However they all 
missed the target, that is the question of the time.  Heisenberg clearly 
stated that indeterminacy relationships do not allow a simultaneous 
measurement of $q$ and $p$, while do not prevent from measuring $q$ and 
$p$ taken in isolation \cite{HEISENBERG}. It is possible to measure, 
with great precision, complementary observables in two different time 
instants: this is not forbidden by Heisenberg's principle.  Later 
on, it is also possible to reconstruct one of the observables at the reference 
time of the other observable, but this is questionable.  In the interval between 
two measurements observables change because time flows.  What 
happened during this interval?  We can reconstruct observables by making 
hypoteses, but we have to remember that these are hypoteses and not 
measurements.

We have to take into account the so--called ``energy--time uncertainty 
relationship''.  As known, time is a $c$--number and therefore it have 
to commute with each operator.  Nevertheless, the relationship exists, 
but it is worth to note its dynamical nature, whereas indeterminacy is 
kinematic \cite{AHARONOV}.  That is, it follows from the evolution of 
the system during the measurement. Bohr had already stated this and he had 
often pointed out the time issue \cite{BOHR28} \cite{BOHR35}, along with 
Landau and Peierls \cite{LANDAU}.  We refer to \cite{LANDAU} in 
which the question is stated in a better way.  The relationship:

\begin{equation}
	\Delta E \Delta t > \hbar
	\label{e:energy}
\end{equation}

\noindent means that we have to consider the system evolution  
during the measurement, that is the difference between the 
measurement result and the state after the measurement.  The 
energy difference between the two states cannot be less than 
$\hbar/\Delta t$.  The energy--time relationship has important 
consequences particularly as regards the momentum measurement and, 
therefore, on double--slit experiment \cite{LANDAU}.

Eq.~(\ref{e:energy}) suggests that, given a certain 
energy, it is possible to construct a state with a huge $\Delta E$ in order 
to obtain a very small $\Delta t$. However, in a recent paper, Margolus and Levitin
\cite{MARGOLUS} give a strict bound that depends on the difference between the average
energy of the system and its ground state energy. Is it a step toward a quantization
of the time?

In addition, if we consider the equation of motion (written with 
Dirac's notation \cite{DIRAC}):

\begin{equation}
	i\hbar \frac{d|Pt>}{dt}=H(t)|Pt>
	\label{e:moto}
\end{equation}

\noindent we can see that $H(t)$ is $i\hbar$ times an operator of 
time--translation.  If the system is closed we can consider $H$ 
constant and equal to the total energy of the system; but if not, if 
energy depends on time, this means that the system is under the action 
of external forces (e.g.  measurement).  The measurement introduce an 
energy exchange that does not follow causality.

Moreover, it is worth to note that a closed system is an 
abstraction.  A real closed system is not observable, without 
introducing energy exchange which would change $H$: therefore that would not 
be a closed system. We can say, by means of Rovelli's words that there is no way to
get information about a system without physically interacting with it for a certain
time \cite{ROVELLI}.

Would you consider it a sophism?  Of course not. We should always bear in mind that 
quantum physics is only an interpreted language we use to speak about 
Nature, though it does not describe Nature itself.  In classical 
physics we made many approximations, which are no longer valid in 
quantum physics. In particular, we can no more neglect time. 
As Heraclitus stated, you cannot plunge your hands twice in the same stream.

\section{Conclusions}
In this paper, the formal character of quantum mechanics is 
emphasized, showing the clear distinction between mathematics and 
nature, words and objects.  As Bohr used to say, physics concerns what 
we can say about nature, by using the language of mathematics.  The 
interpretation of quantum mechanics, that is a correlation between a 
symbol and an object, is only a limit process, because every 
experiment is macroscopic, in order to be reached to human beings.  
Owing to its formal character, quantum theory is subjected to 
G\"{o}del's incompleteness theorems. 

Bohr's complementarity can be a useful tool to investigate time, 
because emphasize that the flowing of time prevent 
the birth of contradictions in nature. Time makes a cut between 
two different, but complementary aspects of reality.  The link 
between complementarity and time requires further investigations.

\section{Acknowledgements}
The author wishes to thank S.~Bergia, F.~Bonsignori, F.~Cannata, and 
F.~Palmonari, for their helpful discussions. 
A special thank to G.~Lolli, for valuable comments on G\"{o}del's theorems.

\section{List of symbols}
\begin{itemize}
	\item $\vdash$: formal deduction;

	\item $\models$: semantic deduction;

	\item $\neg A$: not--A;
	
	\item $\vee$: or;
	
	\item $\wedge$: and.
\end{itemize}

\end{document}